\documentclass[aps,twocolumn,showpacs,floatfix,superscriptaddress,prb]{revtex4-1}

\usepackage{amsmath}
\usepackage{graphicx}
\usepackage{amssymb}
\usepackage{bm}
\usepackage{longtable}
\usepackage{color}
\usepackage{times}
\usepackage{hyperref}
\usepackage[T1]{fontenc}
\usepackage{textcomp}
\usepackage{mathptmx}

\definecolor{darkgreen}{rgb}{0,0.5,0}
\begin{document}

\title{ 
Numerical test of the Cardy-Jacobsen conjecture in the site-diluted Potts
model in three dimensions.
}
\author{L.A.~Fernandez}
\affiliation{Departamento de F\'\i{}sica Te\'orica I, 
  Universidad Complutense, 28040 Madrid, Spain.}
\affiliation{Instituto de Biocomputaci\'on and
  F\'{\i}sica de Sistemas Complejos (BIFI), 50009 Zaragoza, Spain.}

\author{A. Gordillo-Guerrero} 
\affiliation{Departamento de Ingenier\'{\i}a El\'ectrica, 
  Electr\'onica y Autom\'atica,
  Universidad de Extremadura, 10071 Caceres, Spain.}
\affiliation{Instituto de Biocomputaci\'on and
  F\'{\i}sica de Sistemas Complejos (BIFI), 50009 Zaragoza, Spain.}

\author{V. Martin-Mayor}
\affiliation{Departamento de F\'\i{}sica Te\'orica I, 
  Universidad Complutense, 28040 Madrid, Spain.}
\affiliation{Instituto de Biocomputaci\'on and
  F\'{\i}sica de Sistemas Complejos (BIFI), 50009 Zaragoza, Spain.}

\author{J.J. Ruiz-Lorenzo}
\affiliation{Departamento de F\'{\i}sica,
  Universidad de Extremadura, 06071 Badajoz, Spain.}
\affiliation{Instituto de Biocomputaci\'on and
  F\'{\i}sica de Sistemas Complejos (BIFI), 50009 Zaragoza, Spain.}

\date{\today}

\begin{abstract}
We present a microcanonical Monte Carlo simulation of the site-diluted Potts
model in three dimensions with eight internal states, partly carried out in
the citizen supercomputer Ibercivis.  Upon dilution, the pure model's
first-order transition becomes of the second-order at a tricritical point.  We
compute accurately the critical exponents at the tricritical point. As
expected from the Cardy-Jacobsen conjecture, they are compatible with their
Random Field Ising Model counterpart.  The conclusion is further reinforced by
comparison with older data for the Potts model with four states.

\end{abstract}
\pacs{
05.50.+q, 
64.60.De, 
75.40.Mg 
} 
\maketitle 

\section{Introduction}

When two ordered phases compete, even a tiny amount of disorder is
significant. Consider, for instance, the  antiferromagnetic insulator $\rm La
Cu O_4$. A small La $\leftrightarrow$ Sr substitution turns it into a
high-temperature superconductor. Also for colossal magnetoresistance oxides
the importance of the combination of phase coexistence and chemical disorder
has been emphasized.\cite{MANGA} 

These examples suggest a simple, yet general question: {\em which are
  the effects of quenched disorder on systems that undergo a
  first-order phase transition in the ideal limit of a pure sample?}
(quenched disorder models impurities that remain static over
experimental time-scales\cite{GIORGIO}). In fact, this question has
been relevant in a large number of physical contexts. A non-exhaustive
list includes nanoscale ferroelectricity,\cite{XU09} tilt
ordering,\cite{KOI08} ferroelectric thin films,\cite{HU07,SCO05}
random block copolymers,\cite{WES05} ferroelectric
nanodisks,\cite{NAU04} topological phases in correlated electron
systems,\cite{FRE04} effects of multiplicative noise on electronic RLC
circuits\cite{BER03} and surface waves.\cite{BER03,RES02}

Unfortunately, only for two spatial dimensions ($D\!=\!2$) we have a
good understanding of the effects of quenched disorder on
phase-coexistence: the slightest concentration of impurities switches
the transition from first-order to
second-order.\cite{Aize89,Cardy97,UNIVERSALIDAD2D}

In $D\!=\!3$ we lack a general description. One should consider two different
possibilities: disorder may couple either to the order parameter, as in the
Random Field Ising Model (RFIM),\cite{Natherman,Belanger} or it may couple to
the energy, as in the disordered Potts model.\cite{WU} In both cases,
quenched disorder is unreasonably efficient at softening the transition. It
has been surprisingly difficult to show that the transition actually remains
of the first-order for \emph{some} range of impurity
concentration.\cite{Bricmont87,POTTS3D,Igloi2006}

\begin{figure}
\begin{center}
  \includegraphics[height=0.7\columnwidth,angle=0]{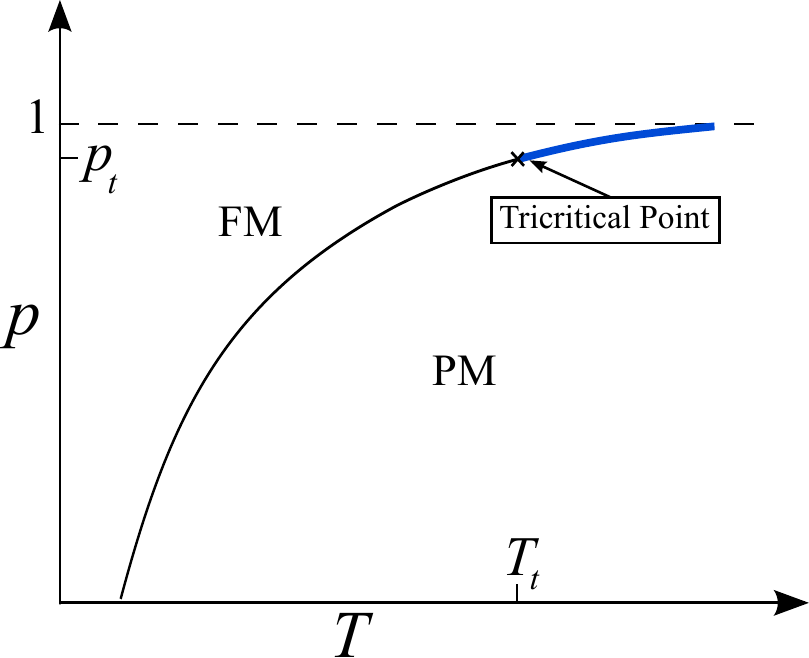}
  \caption{(Color online) Phase diagram of the three dimensional diluted Potts
    model for $Q \ge 3$. For small dilutions we have a first order phase
    transition line which ends up in a tricritical point [at $(T_t,p_t)$], and
    below this tricritical point, the phase transition line continues as a
    second order one. PM and FM denote a paramagnetic and ferromagnetic phase,
    respectively.}
\label{fig_phase_diagram}
\end{center}
\end{figure}

Actually, the Cardy and Jacobsen conjecture relates the two types of disorder
by means of a mapping between the RFIM and the disordered Potts
model.\cite{Cardy97} The conjecture reads as follows. Consider a ferromagnetic
system undergoing a first order phase transition for a pure
sample.\cite{Leuzzi} Let $T$ be the temperature while $p$ is the concentration
of magnetic sites (see the generic phase diagram in
Fig.~\ref{fig_phase_diagram}).  A transition line, $T_\mathrm{c}(p)$ separates
the ferromagnetic and the paramagnetic phases in the $(T,p)$ plane.  In
$D\!=\!3$ a critical concentration is expected to exist, $1>p_\mathrm{t}>0$,
such that the phase transition is of the first-order for $p>p_\mathrm{t}$ and
of the second order for $p<p_\mathrm{t}$ (at $p_\mathrm{t}$ one has a {\em
  tricritical point}). When $p$ approaches $p_\mathrm{t}$ from above, the
latent-heat and the surface tension vanish while the correlation-length
diverges. The corresponding critical exponents can be obtained from those of
the RFIM (see below).

However, the Cardy-Jacobsen mapping relates two problems unsolved in
$D\!=\!3$. In particular, the RFIM (the supposedly well-known partner in the
conjecture) suffers from severe inconsistencies between analytical,
experimental and numerical work. On the experimental side, mutually
inconsistent results for the correlation-length exponent $\nu$ were
obtained,\cite{SLA99,YE02} due to the uncertainties in the parameterization of
the scattering line shape. Also, the estimate of the anomalous dimension
$\eta$ violates hyperscaling bounds.\cite{SLA99} Numerical determinations of
exponent $\nu$ are scattered on a wide
range,\cite{DAFF-UCM,Ogielski86,Gofman,Newman96,Swift97,Rieger95,Dukovski03,Angles97,Nowak98,Nowak99,Middleton02,Hartmann}
and hyperscaling-violating results have been reported.\cite{Hartmann} The
order-parameter's critical exponent $\beta\sim 0.01$ is so small (yet, see
Ref.~\onlinecite{YE02}), that it has even been conjectured that the transition
could be of the first order.\cite{SOURLAS99,MAIORANO07}

On the other hand, the investigation of the disordered Potts model has been
mostly numerical up to now.  In the conventional approach, one averages over
disorder the free-energy \emph{at fixed temperature}.\cite{GIORGIO} It works
nicely for the second-order part of the critical line
$T_\mathrm{c}(p)$,\cite{Ball00,Chat01,Chat05,Paoluzzi10,Leuzzi11,malakis12} but
the first-order piece is plagued by huge sample-to-sample fluctuations of the
specific-heat or the magnetic susceptibility.\cite{Chat05} Fortunately, these
wild fluctuations can be avoided by averaging over disorder the entropy
obtained from microcanonical Monte Carlo,\cite{VICTORMICRO} \emph{ at fixed
  energy}.\cite{POTTS3D} We investigated in this way the site-diluted Potts
model with $Q=4$ states. A delicate extrapolation to infinite system size
showed that $p_\mathrm{t}<1$. Unfortunately, the relevance of the RFIM
universality class for the tricritical point (the core of Cardy and Jacobsen
conjecture) could not be addressed up to now.

Here we show that the Cardy-Jacobsen conjecture is verified to high numerical
accuracy in the site-diluted Potts model with $Q\!=\!4$  and 8 states. This results
follows from a finite size scaling analysis of old $Q\!=\!4$
data\cite{POTTS3D} and new, extensive Monte Carlo simulations for $Q\!=\!8$,
partly carried out in the Ibercivis citizen
supercomputer.\cite{IBERCIVIS1} Our analysis benefits from a recent
computation of the RFIM critical exponents,\cite{DAFF-UCM} that also exploits
the redefinition of the disorder average.\cite{POTTS3D}

In Section~\ref{CJ} we summarize the main implications of the
Cardy-Jacobsen conjecture and define our specific model. Our
methodology, including details on simulation and statistical analysis,
is presented in Section~\ref{Method}. In Section~\ref{Results} we
present our main numerical evidences for the validity of the
conjecture.  We give our conclusions in
Section~\ref{Conclusions}. Finally in the appendix we describe how the
Control Variates technique improves the determination of some
important quantities.

\section{The Cardy-Jacobsen Conjecture}
\label{CJ}

Specifically, we consider the $D\!=\!3$ site-diluted Potts model with $Q$
internal states.\cite{WU} The spins, $\sigma_i\!=\!1,\ldots,Q$, occupy the
nodes of a cubic lattice of linear size $L$, with periodic boundary
conditions. Each spin interacts with its nearest neighbors through the
Hamiltonian
\begin{equation}\label{def:Hspin}
{\cal H}^\mathrm{spin}=-\sum_{\langle i,j \rangle}\epsilon_i \epsilon_j
\delta_{\sigma_i \sigma_j}\,.
\end{equation}
The quenched randomness is represented by the occupation variables
$\epsilon_i=0,1$ ($\epsilon_i=1$ means that the $i$-th spin is present). We
choose each $\epsilon_i$ independently, setting $\epsilon_i=1$ with
probability $p$. Each specific disorder realization is called a
\emph{sample}. The pure system is recovered for $p\!=\!1$, where it undergoes
a, generally regarded as very strong, first-order phase transition for
$Q\geq3$.\cite{Chat05,VICTORMICRO} We show in Fig.~\ref{fig_phase_diagram} the full phase diagram of this model.

The Cardy and Jacobsen mapping relates the large-$Q$ limit of the disordered
Potts model to the RFIM.\cite{Cardy97}  At the tricritical point
$p_\text{t}$ of the Potts model, we encounter three relevant scaling fields
(see, e.g., Ref. \onlinecite{VICTORAMIT}). The dilution field lies along the critical line
$T_\mathrm{c}(p)$. We name its scaling dimension $y_p$. The thermal scaling
field has dimension $y_T$, and is responsible for the ferromagnetic transition
when varying the temperature. Finally, the magnetic scaling field is related
to an external magnetic field in Eq.~\eqref{def:Hspin}. The mapping to the
RFIM is
\begin{align}
\label{nu-p}
y_p &= y^\mathrm{RFIM}_{h_R/J} = \frac{1}{\nu^\mathrm{RFIM}} \,, \\
\label{nu-T}
y_T &= y^\mathrm{RFIM}_{H}- \theta =\frac{1}{2}(D-\theta+2-\eta^\mathrm{RFIM})\,,
\end{align}
where $\nu^\mathrm{RFIM}$ is the correlation-length exponent,\cite{FN2}
$\eta^\mathrm{RFIM}$ is the anomalous dimension, while $\theta$ is the
hyperscaling-violations exponent.\cite{Natherman} Furthermore, the exponent
of the surface tension $\mu$ verifies a modified Widom law:
$\mu=D-\theta-1$. Cardy and Jacobsen predicted as well that, upon
approaching the tricritical point $p_t$, the latent heat in the diluted Potts
model vanishes with the same exponent $\beta^\mathrm{RFIM}$ that rules the
vanishing of the order parameter in the RFIM.

\section{Methodology}
\label{Method}

\subsection{The Microcanonical Ensemble}

For the simulation of the model described by Eq.~\eqref{def:Hspin} we
have used an extended microcanical method which is
suitable to study the first order part of the transition line.\cite{VICTORMICRO}

We will briefly review the main facts of this simulation approach.
Using a mechanical analogy, each spin is
complemented with a \emph{conjugated momenta}. The total energy is the sum of
a kinetic term, $\cal K$ (the halved sum of the squared momenta) and the
potential energy, namely the spin Hamiltonian of Eq.~\eqref{def:Hspin}.

We consider the microcanonical ensemble, where the energy (kinetic
plus potential) is kept fixed to the total value $N e$, where
$N=\sum_i \epsilon_i$ is the total number of spins. The momenta can be
explicitly integrated out. The entropy density $s(e)$ and the
microcanonical weight $\omega(e,N;\{\sigma_i\})$ turn out to be
\begin{eqnarray}
\exp[N s(e,N)] &=&\displaystyle \frac{(2\pi N)^{\frac{N}{2}}}{N \Gamma(N/2)}
\sum_{\{\sigma_i\}}
\omega(e,N;\{\sigma_i\})\,,\label{MICRO1}\\ 
\omega(e,N;\{\sigma_i\}) &=&
\left(\frac{\cal K}{N}\right)^{\frac{N-2}{2}}
\theta\big({\cal K}\big)\,,\label{MICRO2}\\ 
{\cal K}&=& N e- {\cal H}^\mathrm{spin}\,.
\end{eqnarray}
The role of the Heaviside step function in Eq.~\eqref{MICRO2} is preventing
the kinetic energy from becoming negative.

The Monte Carlo simulation of the weight in Eq.~\eqref{MICRO1} is
straightforward.  Both Metropolis and cluster methods are feasible and
efficient.\cite{VICTORMICRO,POTTS3D} In the present work we have used
the Swendsen-Wang algorithm\cite{VICTORMICRO} (see
Ref.~\onlinecite{POTTS3D} for implementation details). One obtains in this
way mean-values at fixed $e$ that will be denoted
$\langle(\cdots)\rangle_e$.

A particularly important mean-value comes from the Fluctuation-Dissipation
relation
\begin{equation}\label{SECOND-LAW}
\frac{\mathrm{d} s}{\mathrm{d}e}=\langle  \hat{\beta}\rangle_e\,,
\end{equation}
where 
\begin{equation}
 \hat{\beta}=\frac{N-2}{Ne-{\cal H}^\mathrm{spin}}\,.
 \label{beta_micro}
\end{equation}
On the view of Eq.~\eqref{SECOND-LAW}, it might be inspiring to think
of $\langle \hat{\beta}\rangle_e$ as the inverse-temperature
corresponding to energy density $e$. The connection between the
canonical and the microcanonical ensembles is discussed in
Ref.~\onlinecite{FSSMICRO}. Finally, our main observable will be $\beta(e)$,
defined as
\begin{equation}
\beta(e)=\overline{\langle  \hat{\beta}\rangle_e}\,.
\end{equation}
where the overline stands for the disorder-average as computed at
fixed $e$.

\subsection{The Maxwell construction}\label{sect:the-maxwell-construction}

\begin{figure}
\begin{center}
  \includegraphics[height=\columnwidth,angle=270]{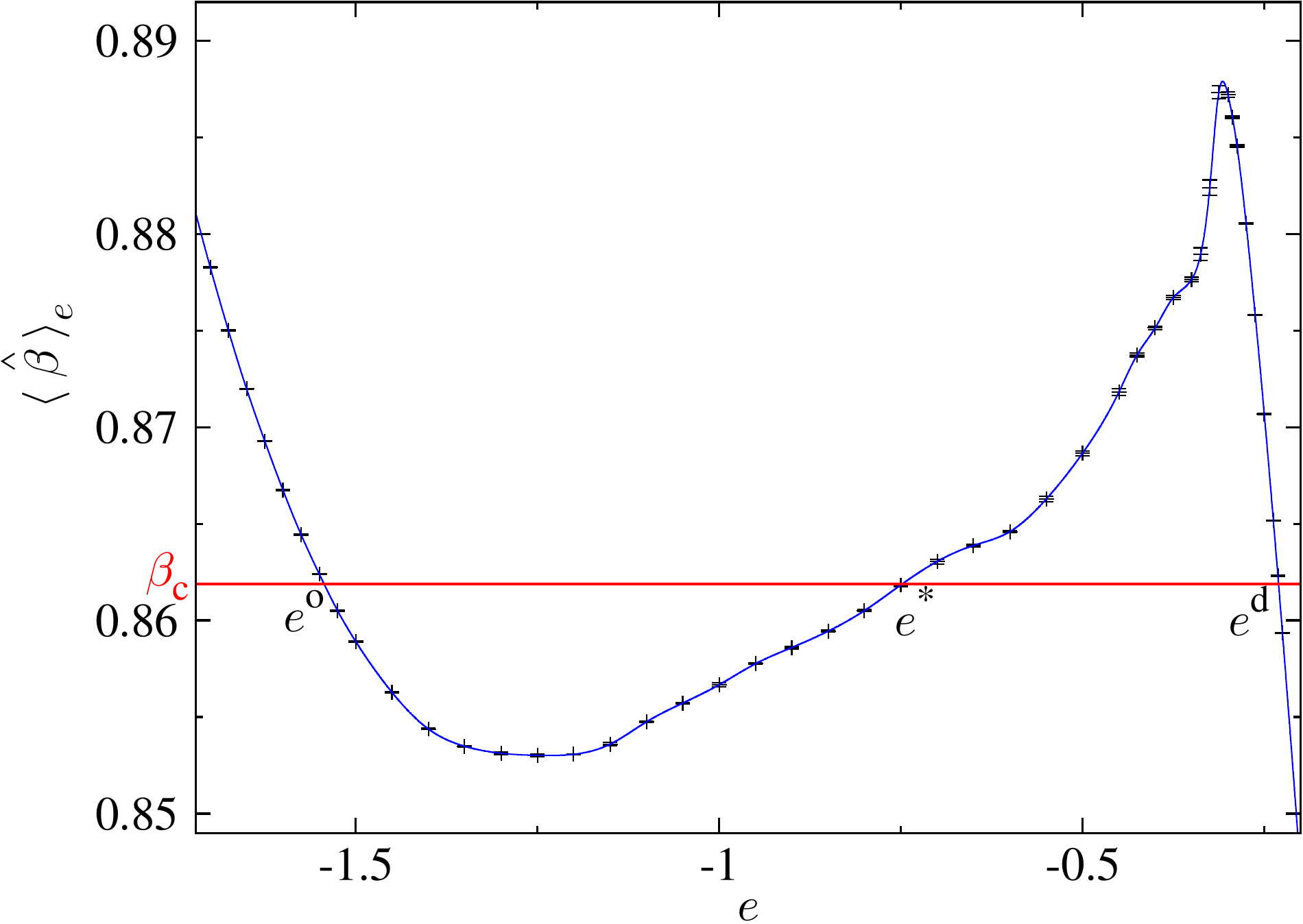}
  \caption{(Color online) Example of Maxwell construction (data from a single sample of a
    $Q=8$ Potts model in three dimensions, with $L=24$ and $p=0.95$). The
    horizontal line corresponds to the inverse critical temperature, obtained
    through Maxwell's equal-area rule, Eq.~\eqref{MAXWELL}. Consider the
    region limited by the horizontal line $\beta=\beta_\mathrm{c}$ and
    the curve $\langle\hat{\beta}\rangle_e$. The (negatively signed) area
     in the region $e^\mathrm{o}< e <e^*$ equals the absolute value of the
     (positively signed) area in the region $e^*<e<e^\mathrm{d}$.
}
\label{fig_maxwell}
\end{center}
\end{figure}

A standard way of studying phase-coexistence in a microcanonical
setting is the Maxwell construction.  This allows to compute from the
curve $\beta(e)$ several important magnitudes: the critical inverse
temperature $\beta_\mathrm{c}$, the energies of the two coexisting
phases and the surface tension. Furthermore, one may apply the very
same method to the sample dependent $\langle
\hat{\beta}_{\{\epsilon_i\}}\rangle_e$, as shown in
Fig.~\ref{fig_maxwell}.  We follow the numerical methods described in
Refs.~\onlinecite{VICTORMICRO} and \onlinecite{POTTS3D}. We briefly
summarize them now, for the sake of completeness.

Consider the equation
\begin{equation}
\beta(e)=\beta\,,\quad \text{or (single sample) } \langle \hat{\beta}_{\{\epsilon_i\}}\rangle_e=\beta\,.\label{eq:gap}
\end{equation}
In normal situations, $\beta(e)$ is monotonically decreasing with $e$, so that
Eq.~\eqref{eq:gap} has a unique solution. However, at phase-coexistence
$\beta(e)$ is no longer monotonically decreasing, see Fig.~\ref{fig_maxwell}.
Therefore, Eq.~\eqref{eq:gap} has three important solutions, named $e^\mathrm{o}$, $e^*$, and $e^\mathrm{d}$
($e^\mathrm{o}< e^* <e^\mathrm{d}$):
\begin{itemize}
\item The rightmost root of~\eqref{eq:gap}, $e_{L,\beta}^\text{d}$,
  corresponds to the ``disordered phase''.
\item The leftmost root of~\eqref{eq:gap}, $e_{L,\beta}^\text{o}$, 
corresponds to the ``ordered phase''.
\item The second rightmost root of~\eqref{eq:gap}, $e_{L,\beta}^*$  is a
  saddle-point among the two phases.
\end{itemize}
Note that these three solutions do depend on $L$, although we
shall not explicitly indicate it unless necessary. 

We compute the
inverse critical temperature $\beta_\mathrm{c}$ from the equal-area rule:
\begin{equation}
0=\int_{e^\text{o}_{\beta_{\text{c}}}}^{e^\text{d}_{\beta_{\text{c}}}}
\text{d}e\, \left(\beta(e) -\beta_{\text{c}}\right)\,,\label{MAXWELL}
\end{equation}
see Fig.~\ref{fig_maxwell}. Note that the $\beta_\mathrm{c}$ computed
from Eq.~\eqref{MAXWELL} does depend on the system size. In fact, in
the thermodynamic limit, Eq.~\eqref{MAXWELL} is a mere consequence of
the continuity of the free-energy density (as a function of
temperature) at the phase transition. In fact, recall that the free
energy density can be expressed in terms of the inverse temperature
and of the internal energy and entropy densities: $f=e - s/\beta$.  Now,
if we recall Eq.~\eqref{SECOND-LAW}, we see that the equality of the
free-energy densities of the ordered and the disordered phases at the
critical temperature can be recast as
\begin{eqnarray}
\beta_\mathrm{c} (e_{\beta_\mathrm{c}}^\mathrm{d} - e_{\beta_\mathrm{c}}^\mathrm{o})&=&
s(e_{\beta_\mathrm{c}}^\mathrm{d})-s(e_{\beta_\mathrm{c}}^\mathrm{o})\,\\
&=& \int_{e^\text{o}_{\beta_{\text{c}}}}^{e^\text{d}_{\beta_{\text{c}}}}
\text{d}e\, \beta(e)\,.
\end{eqnarray}
This textbook reasoning can be extended to the more complicated case
of a finite-system. In fact, it is easy to show, see
Refs.~\onlinecite{VICTORMICRO} and~\onlinecite{JANKEMIC}, that
Eq.~\eqref{MAXWELL} is identical to the criterion of
\emph{equal-height} in the energy histogram.\cite{FSSFO} Such a
finite-system indicator of the critical temperature suffers from
finite-size corrections of order ${\sim 1/L^D}$.\cite{BORGS-KOTECKY}

Once we know $\beta_c$, we may compute the latent-heat as
\begin{equation}
\Delta e = e^\text{d}_{\beta_{\text{c}}}-e^\text{o}_{\beta_{\text{c}}}\,.
\label{latent}
\end{equation}
Finally,  the surface-tension, $\varSigma$, is calculated as
\begin{equation}
\varSigma(L)=\frac{N}{2L^{D-1}} \int_{e^*_{\beta_{\text{c}}^L}}^{e^\text{d}_{\beta_{\text{c}}^L}}
 \mathrm{d}e\,\left( \beta(e) -\beta_\mathrm{c}^L\right)\,.
\label{sigma}
\end{equation}

Note that, in order to compute integrals such as the one in
Eq.~\eqref{MAXWELL}, we interpolate $\beta(e)$ (which is numerically computed
over a grid in the $e$-line), through a cubic spline. Statistical errors are
computed using a jackknife method (see e.g. Ref.~\onlinecite{VICTORMICRO}). In
the case of the sample-averaged $\beta(e)$, the jackknife blocks are formed
from the microcanonical mean-values obtained on the different samples. On the
other hand, when one performs the Maxwell construction for a single sample as
in Fig.~\ref{fig_maxwell}, the jackknife blocks are formed from the Monte
Carlo history.

\begin{figure}
\begin{center}
  \includegraphics[height=\columnwidth,angle=270]{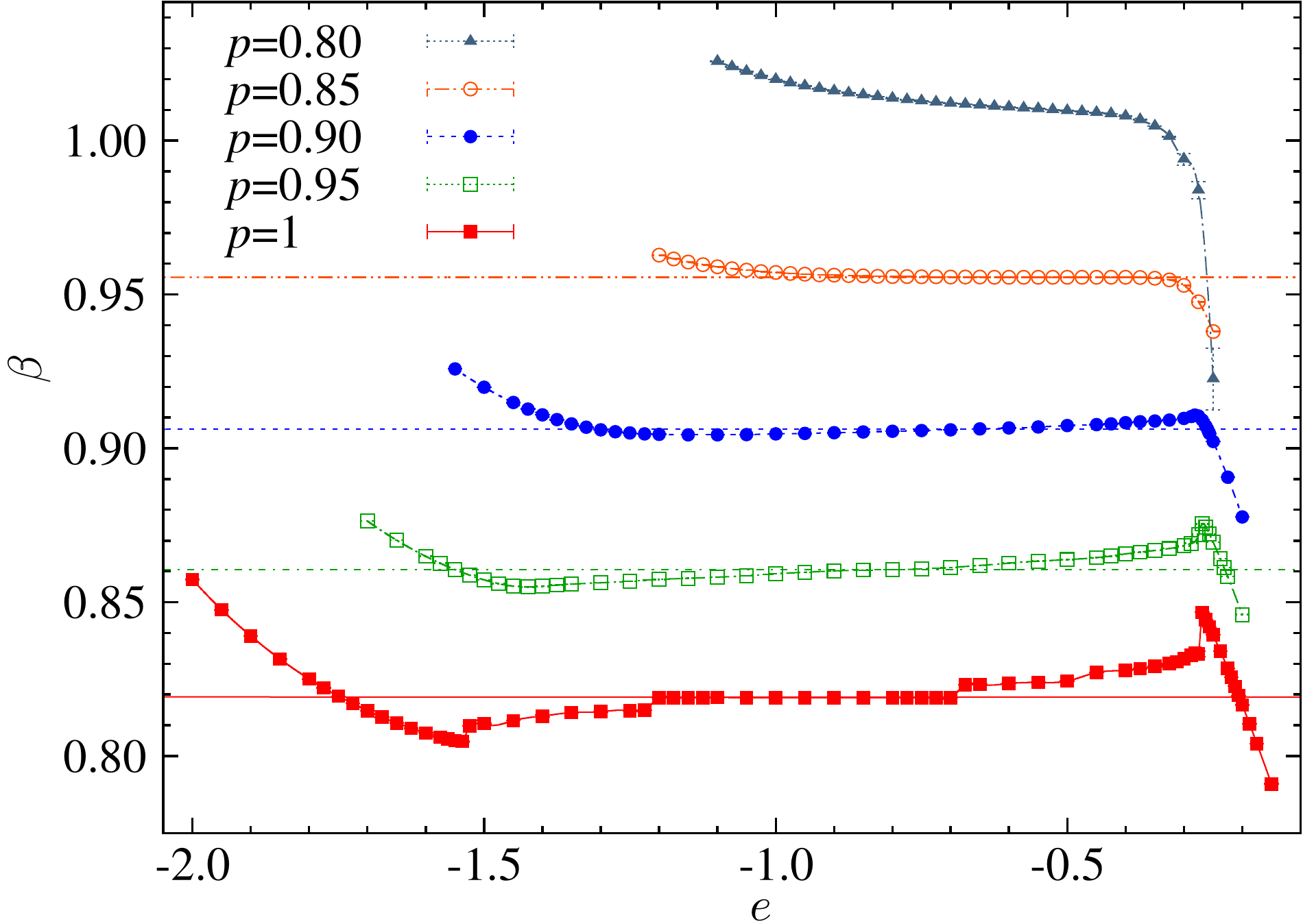}
  \caption{(Color online) Maxwell construction, see Eq.~\eqref{MAXWELL}, as
    obtained for the sample-averaged $\beta(e)$. Data for $L\!=\!48$ and
    several values of the spin concentration. The transition becomes smoother
    as $p$ decreases (from bottom to top). In fact, for $p\!=\!0.8$ the
    Maxwell construction can no longer be done (because the corresponding
    $\beta(e)$ is monotonically decreasing with $e$).}
\label{fig_maxwell_L48}
\end{center}
\end{figure}

It is interesting to compare the curves $\beta(e)$ for fixed $L\!=\!48$, as
the disorder increases (i.e. as $p$ decreases), see
Fig.~\ref{fig_maxwell_L48}.  In the limit of a pure system, $p\!=\!1$,
$\beta(e)$ displays the expected cusps and steps for a system with well
developed geometric and condensation transitions.\cite{FOOT_GEOMETRIC} As
soon as the system becomes disordered, the transition becomes smoother: both
the latent heat, see Eq.~\eqref{latent}, and the surface tension,
Eq.~\eqref{sigma}, are sizably smaller for $p=0.95$ than for $p=1$. This trend
is maintained for decreasing $p$, to the point that the phase transition is
clearly of the second order at $p=0.8$ (for that dilution, $\beta(e)$ is
monotonically decreasing with $e$). We note as well that the curve $\beta(e)$
for $p<1$ is remarkably featureless, specially if compared to its $p=1$
counterpart. Actually, geometric transitions are also found for individual
samples at $p\!=\! 0.95$. However, the energies at which this singular
behavior arise depend on the considered sample, which results in a smooth
averaged $\beta(e)$.

\subsection{Finite Size Scaling near a Tricritical Point}
\label{FSS}

In the following we will discuss some relevant facts about the scaling near a
tricritical point, see,
e.g., Ref.\onlinecite{VICTORAMIT}. Consider some quantity $O$, that, in the
thermodynamic limit, scales as $O^{(L=\infty)}\sim \xi^x$, where $\xi$ is the
correlation length. The Finite Size Scaling (FSS) ansatz, tells us how the same quantity behaves in
a finite system of size $L$. Close to the tricritical point at
$(p_\mathrm{t},T_\mathrm{t}=T_\mathrm{c}(p_\mathrm{t}))$
\begin{equation}\label{FSS-1}
O(L,p_\mathrm{t}+\delta p,T_\text{t}+\delta T)=L^x G(L^{y_T} u_T,L^{y_p} u_p)\,,
\end{equation}
where $G$ is a scaling function, and we have neglected scaling corrections.
As stated in Eqs.~\eqref{nu-p} and ~\eqref{nu-T}, there are two relevant scaling fields, the
thermal field $u_T$ and the disorder field $u_p$.  Both $u_T$ and $u_p$ are
functions of $\delta p$ and $\delta T$, the deviations from the tricritical
point. If we work at $u_T=0$, we should expect that, at linear order,
$\left. u_p\right|_{u_T=0}\propto\delta p\,$. Then the phase transition
is of the second order if $\delta p<0$, and of the first order if $\delta p>0$. 

Our main assumption will be that the Maxwell construction,
see Ref.~\onlinecite{VICTORMICRO} and the previous subsection, enforces the constrain $u_T=0$ to an
accuracy of order ${\cal O}(L^{-D})$ (this expectation is well founded in the
first-order part of the critical line\cite{FSSFO}). Hence,
Eq.~\eqref{FSS-1} simplifies to
\begin{equation}\label{FSS-2}
O(L,p,\text{Maxwell})=L^x \tilde G\big(L^{y_p}(p-p_\mathrm{t})\big)\big(1+ {\cal O}(L^{y_T-D})\big)\,.
\end{equation}
So, the Maxwell construction allows us to employ standard
FSS,\cite{VICTORAMIT} with an effective scaling-corrections exponent
$\omega=D-y_T$.
The combination of Eqs.~\eqref{nu-p} and ~\eqref{nu-T},
standard RFIM scaling relations\cite{Natherman} and the numerical estimates
in Ref.~\onlinecite{DAFF-UCM} yield
$\omega=\theta+\beta^\text{RFIM}/\nu^\text{RFIM}=1.48(2)$.

A further irrelevant scaling field $u_Q=1/\log Q$ with exponent $-\theta$ is
also present.\cite{Cardy97} Numerically, $\theta=1.468(2)$~\cite{DAFF-UCM},
while we expect $\omega=1.48(2)$. These two exponents are so similar that,
given our limited numerical accuracy, we shall not attempt to distinguish
them. However, we remark that one expects a larger amplitude of the
scaling corrections for $Q=4$, what is confirmed by our data (see
Fig.~\ref{fig_omega_nu}).

\subsection{Numerical Simulations and Thermalization Checks}
\label{simulations}

\begin{figure}
\begin{center}
  \includegraphics[height=\columnwidth,angle=270]{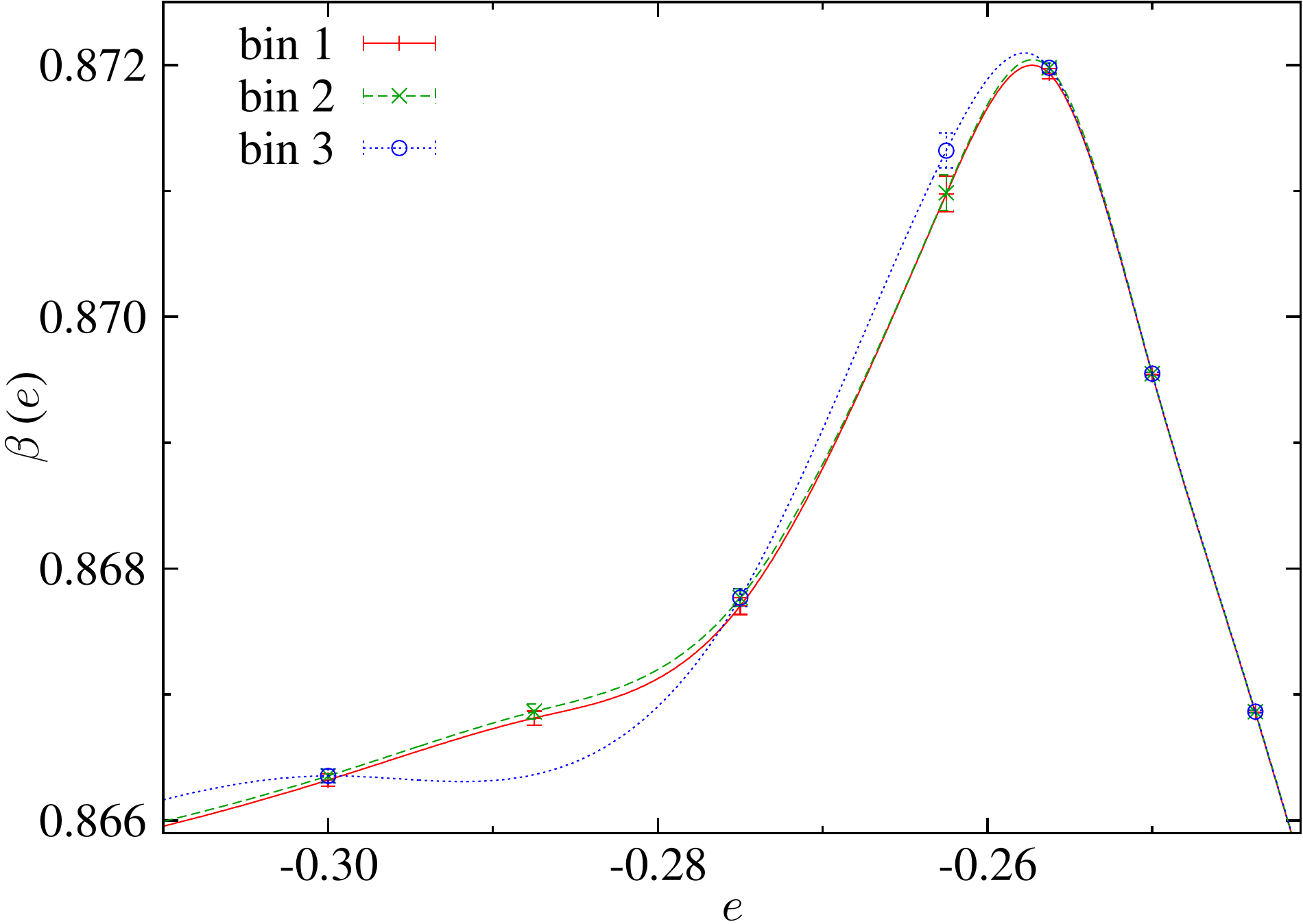}
  \caption{(Color online) In order to ascertain thermalization, we use the standard
    logarithmic data binning (data corresponding to $\beta(e)$, as computed for
    $L=64$, $p=0.95$). Bin 1 was computed from the sample-average of the last
    half of the Monte Carlo history on each sample (bin 2 corresponds to the
    second quarter of the Monte Carlo history, bin 3 to the second eighth, and
    so forth). Statistical compatibility among the different bins is a strong
    thermalization check. Lines are cubic spline-interpolations for each
    bin. In order to demonstrate the importance of having a dense enough
    simulation grid (in particular, close to high curvature regions), the spline
    interpolation in the blue line ignores the data at $e=-0.2875$.}
\label{fig_therma}
\end{center}
\end{figure}

\begin{table}
\centering
\begin{ruledtabular}
\begin{tabular}{c|p{8cm}}
$L$ & \multicolumn{1}{c}{Simulated $p$ values}\\
\hline
12 & 0.65, 0.675, 0.7, 0.725, 0.75, 0.775, 0.8, 0.825, 0.832,        0.85,                0.875,                0.9, 0.925, 0.9375, 0.95\\\hline
16 & 0.65, 0.675, 0.7, 0.725, 0.75, 0.775, 0.8, 0.825,               0.85, 0.854,         0.875,                0.9, 0.925, 0.9375, 0.95\\\hline
24 &              0.7, 0.725, 0.75, 0.775, 0.8, 0.825, 0.832, 0.845, 0.85,                0.875,                0.9, 0.925.\\\hline
32 &                          0.75, 0.775, 0.8, 0.825,               0.85, 0.854, 0.8625, 0.875, 0.886, 0.8875, 0.9, 0.925, 0.9375, 0.95, 0.975.\\\hline
48 &                          0.75, 0.775, 0.8, 0.825,               0.85,        0.8625, 0.875, 0.877, 0.8875, 0.9, 0.925, 0.9375, 0.95.\\\hline
64 &                                       0.8, 0.825,               0.85,       0.86875, 0.875,        0.8875, 0.9, 0.925, 0.9375, 0.95.\\
\end{tabular}
\end{ruledtabular}
\caption{For each of the lattice sizes $L$, we indicate the values of $p$ (the
  concentration of magnetic sites) for which we carried out simulations. We
  shall need to regard the various quantities defined, as continuous functions
  of the density of magnetic sites, $p$. We shall need as well the
  corresponding $p$-derivatives.  As a rule, we have obtained these functions
  of $p$ through a cubic-spline interpolation of the data computed at these
  $p$-values. In fact, some of them were chosen in order to minimize the
  interpolation errors at some particularly important values of $p$, see
  Tables \ref{tab:QUOT-Q8} and \ref{tab:QUOT-Q4}.  Derivatives with
  respect to $p$ were computed simply by derivating the cubic-spline
  interpolating function. The error estimates where obtained through a
  jack-knife (see for instance Ref.~\onlinecite{VICTORAMIT}) over the sample-averages.}
\label{tab:simulations}
\end{table}

We considered concentration values $ 0.65 \leq p\leq 1$ and lattice sizes $12
\leq L \leq 64$. The precise values are indicated in
Table~\ref{tab:simulations}. The $p$-resolution becomes denser close to the
$L$-dependent position of the tricritical point. For all pairs ($p$, $L$) we
simulated 500 samples, with the obvious exception of $p\!=\!1$.

Each sample was simulated on a $e$-grid fine enough to allow for a correct
spline interpolation, see Fig.~\ref{fig_therma}. The simulations at the
different $e$ values were mutually independent. Hence, we faced an embarrassingly
parallel computational problem, suitable for Ibercivis (with a caveat, see
below).

All samples were simulated for the same number of Monte Carlo steps, at every
$e$ value. However, the number of Monte Carlo steps did depend on $e$, as we
explain now. First, we ran all samples at a given $e$-value for a fixed amount
of Swendsen-Wang steps (e.g. $3\times 10^5$ for $L\!=\!64$, or $2\times 10^5$
for $L\!=\!48$), then we assessed thermalization. 

The thermalization check was the standard logarithmic data-binning: for any
given value of $e$, we computed different estimates of the sample-averaged
$\beta(e)$, using disjoint pieces of the Monte Carlo history. On the first
bin, we included only the second-half of the Monte Carlo history (i.e., our
safest data from the point of view of thermalization). The second bin
contained only the second quarter of the Monte Carlo history, etc. We checked
for statistical compatibility, at least, among the first and second bins, see
Fig.~\ref{fig_therma}.  If for a given value of $e$ the thermalization
criterion was not met, the total simulation time was doubled. The procedure
was cycled until convergence was achieved. We note that, for the
concentrations nearest to $p=1$, we encountered strong metastabilities, that
prevented us from simulating $L\!=\!128$ (that could instead be simulated for
$Q\!=\!4$ in Ref.~\onlinecite{POTTS3D}).

The thermalization protocol is not well suited for Ibercivis, because the
simulation of a given sample at some difficult energy may last up to some
days. Yet, Ibercivis relies on volunteers' computers that frequently switch
from on-line to off-line. To minimize the number of unfinished simulations, we have implemented a 
continuity system. It divides every simulation, no matter how long it is,
in small time steps (typically 30 minutes). After every step, consistency checks are performed
and the current system configuration is sent again to the simulation queue. This solved
the problem for relatively long (5-6 hours) simulations but the few more demanding simulations
were completed on local clusters.

Altogether, this work has consumed (the equivalent of) $3\times10^6$ hours of a
single Intel Core2 duo at 2.5 GHz.

We should also mention that we have performed some new, short simulations for
$Q=4$ at $p=0.95$, complementary to those reported in
Ref.~\onlinecite{POTTS3D}. The simulated sizes were $L=24$ and
$L=48$ (128 samples each). Our goal was to improve the accuracy of the
interpolations described below. 

\section{Results}
\label{Results}

To check the Cardy-Jacobsen conjecture we have performed numerical simulations for
$Q=8$, hence further approaching the large-$Q$ limit where the mapping becomes
exact.\cite{Cardy97}

\begin{figure}
\begin{center}
\includegraphics[height=\columnwidth,angle=270]{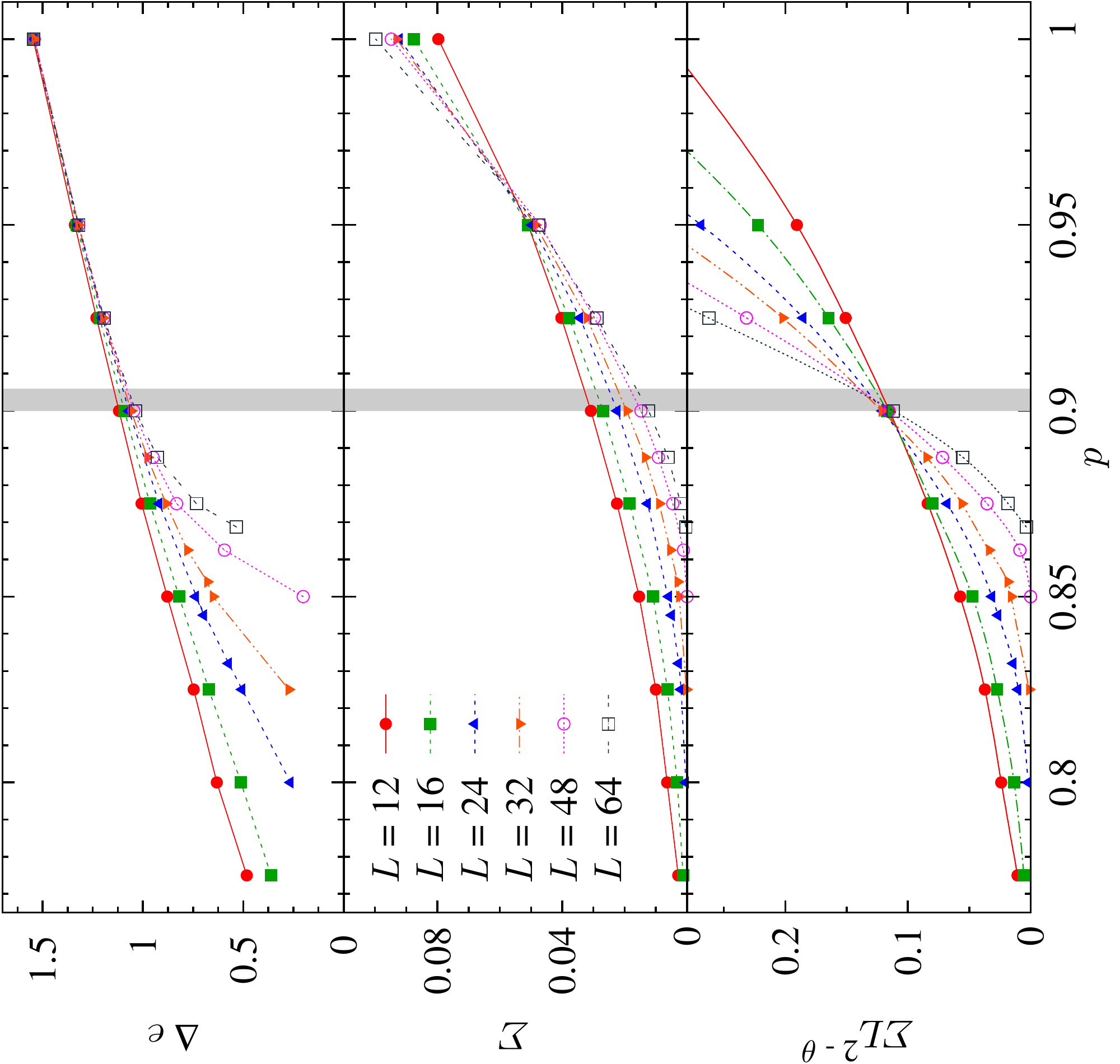}
\caption{(Color online) Latent heat $\Delta e$ ({\bf Top}) and surface tension $\varSigma$
  ({\bf Middle}) as a function spin concentration, $p$, for each lattice size
  (lines are linear interpolations). Lines end at the smallest $p$ that
  allowed to perform the Maxwell construction for each $L$.  {\bf Bottom:}
  Scaled surface tension using $\theta=1.469(20)$~\cite{DAFF-UCM} (the lines
  joining the data are cubic splines). The vertical gray line shows the
  infinite volume extrapolation for $p_{\mathrm t}$.}
\label{fig_latent_tension}
\end{center}
\end{figure}

Consider the $p$ and $L$ evolution of the latent-heat and the surface tension
in Fig.~\ref{fig_latent_tension}.  If $p\!<\!p_\text{t}$ (i.e.  if we are in
the second-order piece of the critical line), both $\Delta e$ and $\varSigma$
vanish in the large-$L$ limit (the two are positive for
$p\!>\!p_\text{t}$). However, for small lattices, both $\Delta e$ and $\varSigma$
decrease gently upon decreasing $p$ which suggests that dilution merely
smoothed the first-order transition.  However, the curve for $\Delta e$
becomes sharper upon increasing $L$. Indeed the Potts-RFIM
mapping\cite{Cardy97} implies $\Delta e \propto (p-p_{\text{t}})^\beta$ with
$\beta\!=\!\beta^{\mathrm{RFIM}}\!\sim\!  0.01$,\cite{Natherman} which is
barely distinguishable from a discontinuous jump. Furthermore, the
$L$-dependent position of the tricritical point $p_\mathrm{t}^L$ (for
instance, the point of sharpest drop of $\Delta e$ in
Fig.~\ref{fig_latent_tension}---top) grows quickly with $L$.  On the view of
the $D\!=\!2$ no-go theorems,\cite{Aize89} one could be afraid that
$p_\mathrm{t}^L\to 1$ for large $L$ also in $D\!=\!3$. We know that this is
not the case,\cite{POTTS3D} but it is clear that a careful scaling analysis is
needed.

Eq.~\eqref{FSS-2} tells us that $L^{2-\theta}\varSigma $ is scale-invariant, and
thus allows to locate the tricritical point (because
$x_\varSigma\!=\!\theta-D+1\!=\!\theta-2$,
$\theta\!=\!1.469(20)$~\cite{DAFF-UCM}). Indeed, in
Fig.~\ref{fig_latent_tension}---bottom, we see that the curves
for system sizes $L_1<L_2$ cross at $p_\mathrm{t}^{L_1,L_2}$:
\begin{equation}\label{eq:crossing}
L_1^{2-\theta}\varSigma(L_1,p_\mathrm{t}^{L_1,L_2})=L_2^{2-\theta}\varSigma(L_2,p_\mathrm{t}^{L_1,L_2})\,,
\end{equation}
($p_\text{t}^{L_1,L_2}\to p_\mathrm{t}$ when $L_1$ diverges). We recall that a
similar method was used recently in a spin-glass
context.\cite{Janus2010corto} There are two main consequences of choosing a
wrong estimate of exponent $\theta$ in Fig.~\ref{fig_latent_tension}---bottom
and Eq.~\eqref{eq:crossing}. First, in the limit of large lattice sizes, the
height of the crossing point diverges (or goes to zero) if $\theta$ is
underestimated (overestimated). Second, the size corrections to the crossing
points are larger for a wrong $\theta$. Specifically,
$p_\mathrm{t}^{L_1,L_2}-p_\text{t} \!=\!{\cal O}(L_1^{-y_p})$. The amplitude
for these scaling corrections cancels only for the exact choice of $\theta$.

The critical exponent for a quantity $O$ is obtained from its quotients at
$p_\mathrm{t}^{L_1,L_2}$:\cite{QUOTIENTS,NIGHTINGALE}
\begin{equation}\label{eq:quo}
\left.\frac{O(L_2)}{O(L_1)}\right|_{p_\mathrm{t}^{L_1,L_2}}=\left(\frac{L_2}{L_1}\right)^{x_O}\,\left[1+A_O\bigg(\frac{1}{L_2^\omega}-\frac{1}{L_1^\omega}\bigg)\right]\,.
\end{equation}
Above, we included only the leading scaling-corrections ($A_O$ is an
amplitude). We use Eq.~\eqref{eq:quo} for the logarithmic $p$-derivative of
$\varSigma$ (scaling dimension $x=y_p$), and for the latent heat (scaling
dimension $x=\beta y_p$, which should be $\beta^\text{RFIM}/\nu^\text{RFIM}$,
according to Cardy and Jacobsen\cite{Cardy97}). Our results are in
Table~\ref{tab:QUOT-Q8} ($Q=8$), and Table~\ref{tab:QUOT-Q4} ($Q=4$). In both
cases we see that the convergence of $p_t^{L_1,L_2}$ to the thermodynamic
limit is very fast. The height of the crossing point seems also stable with
growing sizes.

\begin{table}
\begin{ruledtabular}
\begin{tabular}{lllll}
$(L_1,L_2)$ 
& \multicolumn{1}{c}{$p_t^{L_1,L_2}$} 
& \multicolumn{1}{c}{$y_p$} 
& \multicolumn{1}{c}{$L_1 ^{2-\theta} \varSigma^\text{cross} $} 
& \multicolumn{1}{c}{$\beta y_p$} \\
\hline
(12,16)   & 0.8947(38)(17) &   0.89(23)(2)   & 0.108(5)(3)  &  0.095(9)(5) \\
(12,24)   & 0.8942(16)(15) &   0.82(8)(2)    & 0.107(3)(3)  &  0.075(3)(4) \\
(16,24)   & 0.8939(28)(14) &   0.79(18)(3)   & 0.107(6)(4)  &  0.061(5)(3) \\
(16,32)   & 0.8966(13)(11) &   0.85(13)(3)   & 0.111(3)(4)  &  0.050(2)(2) \\
(24,32)   & 0.8989(28)(10) &   0.94(26)(3)   & 0.118(8)(5)  &  0.035(5)(2) \\
(24,48)   & 0.9031(14)(10) &   0.80(6)(05)   & 0.128(4)(6)  &  0.027(2)(2) \\
(32,48)   & 0.9057(21)(9)  &   0.84(10)(1)   & 0.139(8)(7)  &  0.021(4)(2) \\
(32,64)   & 0.9040(11)(8)  &   0.86(5)(1)    & 0.134(5)(7)  &  0.023(3)(1) \\
(48,64)   & 0.9026(21)(5)  &   0.99(14)(3)   & 0.126(10)(8) &  0.024(6)(1)  \\
\end{tabular}
\end{ruledtabular}
\caption{Quotient-method for $Q=8$. For each pair of lattices $(L_1,L_2)$,
    we extract the crossing point $p_t^{L_!,L_2}$, see
    Eq.~\eqref{eq:crossing}, and the height of the crossing point,
    $\varSigma(L_1,p_t^{L_1,L_2}) L_1 ^{2-\theta}$. The effective critical
    exponents $y_p$ and $\beta y_p$ are obtained using the quotients method,
    Eq.~\eqref{eq:quo}.  For each data, we indicate two error bars. The first
    error is statistical.  The second error is due to the uncertainty in
    $\theta= 1.469(20)$.\cite{DAFF-UCM}}
\label{tab:QUOT-Q8}
\end{table}

\begin{table}
\begin{ruledtabular}
\begin{tabular}{lllll}
$(L_1,L_2)$ 
& \multicolumn{1}{c}{$p_\mathrm{t}^{L_1,L_2}$} 
& \multicolumn{1}{c}{$y_p$} 
& \multicolumn{1}{c}{$L_1 ^{2-\theta} \varSigma^\text{cross} $} 
& \multicolumn{1}{c}{$\beta y_p$}\\
\hline
(16,24)    & 0.9249(30)(8)  &   1.40(46)(3)   & 0.0113(6)(5)   &  0.285(11)(6) \\
(16,32)    & 0.9324(19)(8)  &   1.11(20)(5)   & 0.0125(5)(6)   &  0.230(6)(6) \\
(24,32)    & 0.9400(30)(6)  &   1.22(33)(1)   & 0.0159(12)(8)  &  0.175(12)(4) \\
(24,48)    & 0.9455(19)(9)  &   0.83(8)(3)    & 0.0179(9)(8)   &  0.135(5)(4) \\
(32,48)    & 0.9506(27)(8)  &   0.79(18)(3)   & 0.0215(17)(10) &  0.112(7)(3) \\
(32,64)    & 0.9489(13)(7)  &   0.78(9)(2)    & 0.0206(9)(11)  &  0.095(4)(3) \\
(48,64)    & 0.9473(31)(5)  &   0.92(24)(3)   & 0.0191(25)(12) &  0.070(10)(3) \\
(48,128)   & 0.9491(9)(5)   &   0.77(8)(2)    & 0.0204(10)(13) &  0.048(4)(3) \\
(64,128)   & 0.9497(14)(5)  &   0.71(13)(2)   & 0.0213(17)(14) &  0.038(6)(3)  \\
\end{tabular}
\end{ruledtabular}
\caption{Quotient method for $Q=4$ (data from
  Ref.~\onlinecite{POTTS3D}, improved through control variates and the
  addition of new runs near $p_\mathrm{t}$).  Same notations of
  Table~\ref{tab:QUOT-Q8}.}
\label{tab:QUOT-Q4}
\end{table}

The results in Tables~\ref{tab:QUOT-Q8} and~\ref{tab:QUOT-Q4} need to be
extrapolated to the limit of infinite system sizes. This can be done by
considering leading order scaling corrections, as in Eq.~\eqref{eq:quo}. The
extrapolation greatly improves by imposing to $Q=4$ and $8$ a common
extrapolation and the same scaling-corrections exponent $\omega$, as required
by the Universality predicted in Ref.~\onlinecite{Cardy97}. 

In this way, we obtain $\beta y_p=0.0022(48)(3)$ and $\omega=1.36(8)(1)$ where
the second parenthesis indicates the uncertainty induced by the error in
$\theta$.\cite{DAFF-UCM} The fit quality is assessed through the $\chi^2$
test.  We obtain $\chi^2/\mathrm{dof}=8.5/14$, which is almost too good
(\emph{dof} stands for the number of degrees of freedom of the fit).  Indeed
the probability of getting such a low value of $\chi^2$ with 14 degrees of
freedom is only $14\%$. We note as well that $\beta y_p=0.0022(48)(3)$ is only
barely compatible with the best RFIM estimate $\beta
y_p=0.0119(4)$,\cite{DAFF-UCM} (since the discrepancy is as large as two
standard deviations).

\begin{figure}
\begin{center}
\includegraphics[height=\columnwidth,angle=270]{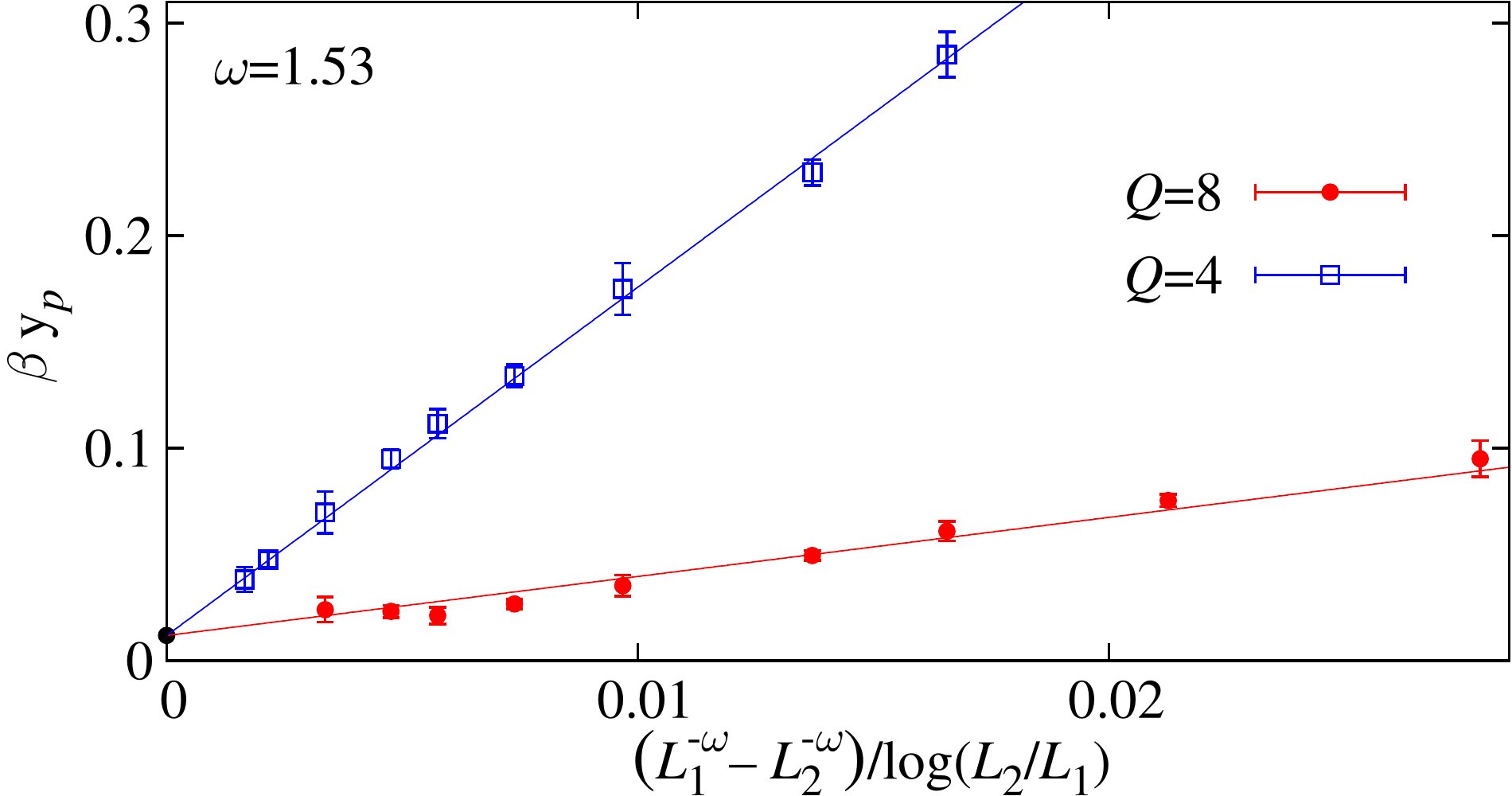}
\caption{(Color online) Determination of the scaling correction exponent
  $\omega$, from the size-dependent effective exponent $\beta y_p$ for the
  vanishing latent-heat at the tricritical point as computed with the
  quotients method, see Tables ~\ref{tab:QUOT-Q8} and~\ref{tab:QUOT-Q4}. A
  common extrapolation $\beta y_p\!=\!0.0119(4)$~\cite{DAFF-UCM} is imposed in
  the joint-fit for the $Q\!=\!4$ and $Q\!=\!8$ data. The figure of merit
  $\chi^2=14/15$ is computed with the full covariance matrix.}
\label{fig_omega_nu}
\end{center}
\end{figure}

At this point, we can try to \emph{disproof} universality. We make the
\emph{assumption} that $\beta y_p$ takes exactly the RFIM value, redo the fit
and see the outcome of the $\chi^2$ test. This second fit, with $\beta y_p$
fixed to $0.0119$, turns out to be perfectly reasonable
($\chi^2/\text{dof}=14/15$, see Fig.~\ref{fig_omega_nu}).  Hence we conclude
that our data set \emph{is} statistically compatible with universality.

A second, unexpected bonus of fixing $\beta y_p$ in the fit to the RFIM value,
is a remarkable increase in the accuracy of $\omega=1.53(5)(3)$, in excellent
agreement with our expected $\omega=1.48(2)$ (remember that
$\omega=D-y_T=\theta+\beta^\text{RFIM}/\nu^\text{RFIM}$, see Sec. \ref{FSS}).
Furthermore, from this value of $\omega$, we obtain $y_T=D-\omega=1.47(8)$, in
nice agreement with the large-$Q$ computation $y_T=1.49(9)$.\cite{Igloi2006}

Following the same approach for $y_p$, which is expected to coincide with
$1/\nu^\text{RFIM}$, we obtain $y_p=0.775(46)(1)$ if we impose $\omega=1.36$
(we get $y_p=0.779(41)(1)$ by taking $\omega=1.53$).  Both fits are fair
($\chi^2/\text{dof}=13.6/15$ and $\chi^2/\text{dof}=13.2/15$,
respectively). 

Our $y_p$ is in the lower range of previous numerical and analytical
estimates: $0.73\leq 1/\nu^\text{RFIM}\leq
1.12$.\cite{DAFF-UCM,Ogielski86,Gofman,Newman96,Swift97,Rieger95,Dukovski03,Angles97,Nowak98,Nowak99,Middleton02,Hartmann,NPFRG}
Hyperscaling and our $y_p$ implies a slightly positive specific heat
exponent $\alpha=(2y_p-D+\theta)/y_p = 0.03(10)$, in agreement with
experimental claims of a (possibly logarithmic)
divergence.\cite{EXPO-ALPHA-BELANGER} We warn however that severe
hyperscaling violations [namely $\alpha = -0.63(7)$] have been
reported in numerical work.\cite{Hartmann}

One may compute as well the exponent $\theta$, by fitting
$\varSigma(L,p_\text{t}^{L,2L})=A_Q L^\theta(1+ B_Q L^{-\omega})$ (only the
amplitudes $A_Q$ and $B_Q$ are $Q$-dependent on the fit). Taking
$\omega=1.5(1)$, we obtain $\theta=1.52(11)(2)$, with an acceptable fit
($\chi^2/\mathrm{dof}=4.9/3$). The result is compatible with, but less
accurate than, the latest RFIM result $\theta=1.469(20)$.\cite{DAFF-UCM}

\section{Conclusions}~\label{Conclusions}

In summary, we have presented a finite-size scaling analysis of the
tricritical point of the site-diluted Potts model in three dimensions for
$Q=4$ and $8$ internal states. By considering leading-order scaling
corrections we were able to show that the relevant Universality class for the
tricritical point is the one of the RFIM. To our knowledge, this is the first
verification of the Cardy-Jacobsen conjecture.\cite{Cardy97}  

Three technical ingredients were crucial to obtain this achievement:
the use of the microcanonical Monte Carlo,\cite{VICTORMICRO} a new
definition of the disorder average,\cite{POTTS3D} and the use of the
citizen supercomputer Ibercivis.\cite{IBERCIVIS1}

\section{Acknowledgements}

We have been partly supported through Research Contracts
Nos. FIS2009-12648-C03 and FIS2010-16587 (MICINN), GR10158 (Junta de
Extremadura), ACCVII-08 (UEX), and from UCM-Banco de Santander.  We thank
Ibercivis for the equivalent of $3\times 10^6$ CPU hours.  The simulations
were completed in the clusters Terminus (BIFI) and Horus (U. Extremadura). We also thank N.~G. Fytas for a careful reading of the manuscript.

\appendix
\section{Control variates}
\label{analysis}

The statistical quality of data may sometimes be significantly increased by means
of a very simple trick, named \emph{control variates} (see e.g.
Ref.~\onlinecite{CV-UCM}).

\begin{figure}[b]
\begin{center}
  \includegraphics[height=\columnwidth, angle=270]{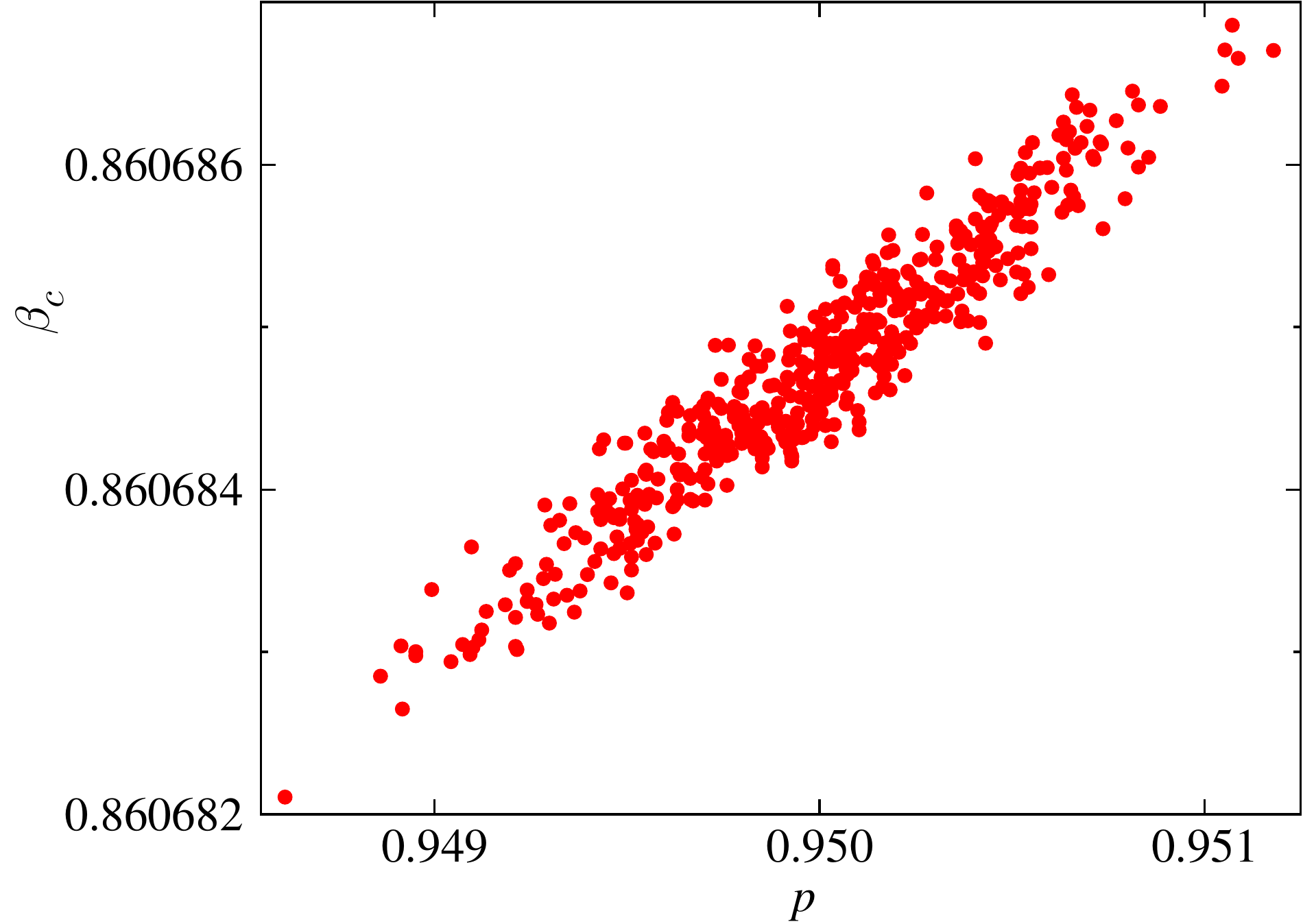}
  \caption{(Color online) Scatter plot of each sample's inverse critical temperature vs. the
    concentration of magnetic sites, $\sum_i\epsilon_i/V$. Data for 500
    samples of $L=64$ and  $p=0.95$. The correlation coefficient that gives the
    optimal coupling to the control variate, see Eq.~\eqref{alpha_optimal}, is
    $\alpha^*=0.956$.}
\label{fig_CVcorr}
\end{center}
\end{figure}

In short, we want to improve our estimation of a stochastic variable $A$
through its correlations with another random variable $B$ ($B$ is named a
control variate).  If $\overline{B}=0$ and $\hat A=A+\alpha B$, then the
expectation value does not change: $\overline{\hat A} =\overline{A}$.
However, depending on the arbitrary election of $\alpha$, we can get
$\text{var}( \hat A )<\text{var}( A )$.  The $\alpha$ election minimizing the
variance $\text{var}( \hat A )$ is
\begin{equation}
\alpha^*=\frac{\text{cov}(A,B)}{\sqrt{\text{var}(A)\text{var}(B)}}\,,
\label{alpha_optimal}
\end{equation}
which coincides with the correlation coefficient $r_{AB}$. The optimal
variance is 
\begin{equation}
\text{var}( \hat A^* )=\text{var}( A)\, (1- r^2_{AB})\,.
\end{equation} 
Hence, the stronger the statistical correlation (or anticorrelation) between
$A$ and $B$, the more effective the control variate is.

\begin{figure}
  \includegraphics[height=\columnwidth, angle=270]{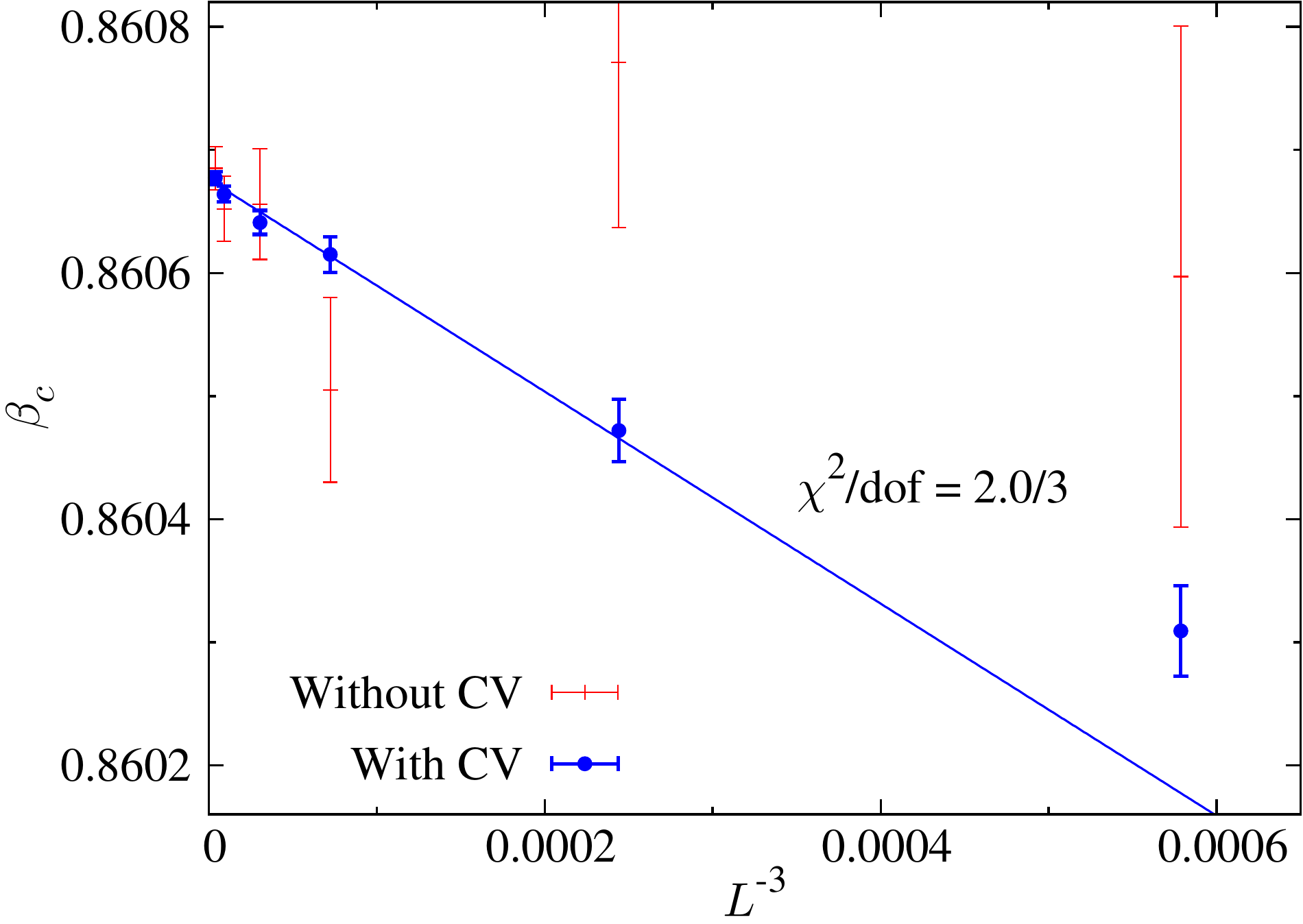}
  \caption{(Color online) Inverse critical temperature $\beta_{\mathrm{c},L}$
    as a function of an inverse lattice volume, $1/L^3$. Data obtained for
    $p=0.95$.  The error reduction obtained with control variates is
    significant (blue points).  In the linear fit we considered only data with
    $L\ge 16$ and improved through control variates.}
\label{fig_CVbeta0.95}
\end{figure}

\begin{figure}
  \includegraphics[height=\columnwidth, angle=270]{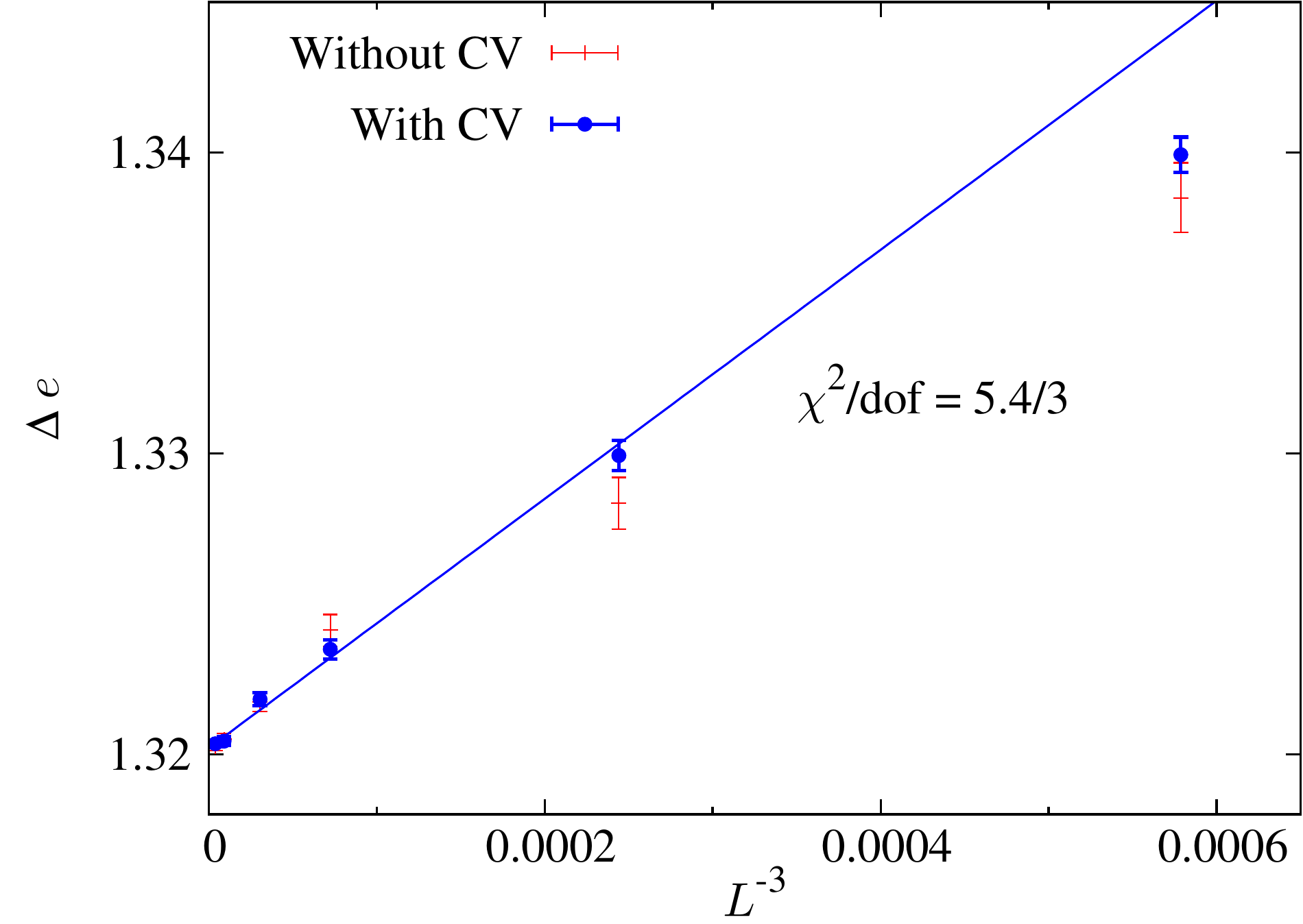}
  \caption{(Color online) The size dependent latent-heat for $p=0.95$, as a function of
the inverse lattice volume. The linear fit includes only data with  $L\geq 16$
that were improved through control variates.}
\label{fig_CVlatent0.95}
\end{figure}

In our case, a rather obvious control variate is
\begin{equation}
B=\frac{1}{V}\sum_i \epsilon_i \ -\ p\,,
\end{equation}
namely the difference among the real and the nominal concentrations of
magnetic sites. It is clear that the disorder average $\overline{B}$
vanishes. We will employ $B$ to improve the determination of the
sample-averaged $\beta(e)$. Note that, although the value of $B$ does
not depend on the considered energy (it is fixed by the
$\{\epsilon_i\}$), its correlation coefficient with
$\langle\hat\beta\rangle_e$ needs to be computed for all energies in
the $e$-grid.

$B$ is extremely effective as a control variate for the computation of the
inverse critical temperature $\beta_\mathrm{c}$, as suggested from
Fig.~\ref{fig_CVcorr}. The correlation coefficient in that plot is so high,
0.956, that the expected error reduction factor is 3.4.  However, the alert
reader will note that this is a hasty conclusion. In fact, the
$\beta_\text{c}$ obtained from $\beta(e)$ is not exactly the average of the
inverse critical temperatures found for each sample. The reason for this
non-linearity in the Maxwell rule, see Eq.~\eqref{MAXWELL}, is that the
energies $e^\mathrm{d,o}$ are not the same for $\beta(e)$ and for the
$\langle\hat\beta\rangle_e$ in a given sample. Yet, the dependency on
$e^\mathrm{d,o}$ of the integral in Eq.~\eqref{MAXWELL} is extremely weak
[recall the stationarity condition with respect to $e$ in Eq.~\eqref{eq:gap}].

In fact, the correct computation with $\beta(e)$ does show a significant error
reduction, see Fig.~\ref{fig_CVbeta0.95}, close to the factor 3.4 anticipated
by the naive analysis in Fig.~\ref{fig_CVcorr}. We note in
Fig.~\ref{fig_CVlatent0.95} an equally significant reduction of the
statistical errors for the latent heat. Therefore, our computation of these
quantities, obtained with only 500 samples, has been made equivalent to a
5000-samples computation. This is a remarkable reward for such a simple
analysis.


\begin{thebibliography}{99}
\bibitem{Leuzzi} One may fit spin-glass interactions into the same scheme, see Ref.~\onlinecite{Paoluzzi10,Leuzzi11}.
\bibitem{FN2} Remember that the parameter that plays the role of the
  reduced temperature in the RFIM is $\tau\simeq
  h_\mathrm{R}/J$,\cite{Natherman}. In the Cardy-Jacobsen mapping, the
  disorder strength in the Potts model, $w$, corresponds exactly with
  $\tau$.\cite{Cardy97} Hence, $y_p=1/\nu^\mathrm{RFIM}$. However,
  keep in mind that this relation was obtained by reasoning in the
  limit $Q\gg 1$.
\end{thebibliography}
\end{document}